%% file: bcast.tex
\documentclass[journal,letterpaper,twoside]{IEEEtran}

\input{quantum}

\usepackage{graphicx}
\usepackage{color}
\usepackage[T1]{fontenc}

\newtheorem{thm}{Theorem}
\newtheorem{lemma}{Lemma}

\newcommand{\aeq}{\approx_{(a)}}

\begin{document}

\title{A father protocol for quantum broadcast channels}
\author{Fr\'ed\'eric Dupuis, Patrick Hayden and Ke Li
\thanks{F. Dupuis is with the Universit\'e de Montr\'eal and McGill University. email: {\tt dupuisf@iro.umontreal.ca}}
\thanks{P. Hayden is with McGill University, email: {\tt patrick@cs.mcgill.ca}}
\thanks{K. Li is with the University of Science and Technology of China, email: {\tt leeke@mail.ustc.edu.cn}}}

\maketitle

\IEEEpeerreviewmaketitle

\begin{abstract}
A new protocol for quantum broadcast channels based on the fully quantum Slepian-Wolf protocol is presented. The protocol yields an achievable rate region for entanglement-assisted transmission of quantum information through a quantum broadcast channel that can be considered the quantum analogue of Marton's region for classical broadcast channels. The protocol can be adapted to yield achievable rate regions for unassisted quantum communication and for entanglement-assisted classical communication; in the case of unassisted transmission, the region we obtain has no independent constraint on the sum rate, only on the individual transmission rates. Regularized versions of all three rate regions are provably optimal.
\end{abstract}


\section{Introduction}\label{sec:intro}

\PARstart{D}{iscrete} memoryless broadcast channels are channels with one sender and multiple receivers, modelled using a probability transition matrix $p(y_1, \ldots, y_n|x)$. There are many natural tasks that one may want to perform using these channels, such as sending common messages to all the users, sending separate information to each user, sending data to each user privately, or some combination of these tasks. Here we shall focus only on sending separate data, and most of our discussions will only involve channels with two receivers.

These channels were first introduced by Cover in \cite{cover1}, where he suggested that it may be possible to use them more efficiently than by timesharing between the different users. Since then, several results concerning broadcast channels have been found, such as the capacity of degraded broadcast channels (see, for example, \cite{coverthomas}).

The best known achievable rate region for general classical broadcast channels is due to Marton \cite{marton}: given a probability distribution $p(x, u_1, u_2) = p(u_1, u_2)p(x|u_1,u_2)$, the following rate region is achievable for the general two-user broadcast channel $p(y_1, y_2|x)$:
\begin{equation}
\label{eqn:marton}
\begin{split}
0 \leqslant R_1 &\leqslant I(U_1; Y_1)\\
0 \leqslant R_2 &\leqslant I(U_2; Y_2)\\
R_1 + R_2 &\leqslant I(U_1; Y_1) + I(U_2; Y_2) - I(U_1; U_2)
\end{split}
\end{equation}
It is conjectured that this characterizes the capacity region of general broadcast channels, but despite considerable efforts, no one has been able to prove a converse theorem.

The quantum generalization of broadcast channels was first studied in \cite{allahverdyan-saakian} and \cite{broadcast} as part of a recent effort to develop a network quantum information theory \cite{horo1,horo2,YDH05,LOW06,netcoding1,winter-qmac,klimovitch-qmac,SVW05,bosonic}. In \cite{broadcast}, the authors derived three classes of results, the first one about channels with a classical input and quantum outputs, the second one about sending a common classical message while sending quantum information to one receiver, and the third about sending qubits to one receiver while establishing a GHZ state with the two receivers.

In this paper, we study quantum broadcast channels using a different approach. Over the past few years, several results in quantum Shannon theory have been unified and simplified by the introduction of the mother and father protocols \cite{mother-father} and, more recently, by the fully quantum Slepian-Wolf (FQSW) protocol \cite{FQSW} \cite{triangle}. A whole array of results, including the mother and the father, can be easily derived from the FQSW protocol, such as the quantum reverse Shannon theorem \cite{reverse-shannon}, the Lloyd-Shor-Devetak (LSD) theorem \cite{lsd1} \cite{lsd2} \cite{lsd3}, one-way entanglement distillation \cite{DW05}, and distributed compression \cite{FQSW}. The results presented here are of the same flavour: we will derive a new coding theorem for general quantum broadcast channels using the FQSW theorem. The new protocol plays the role of a father protocol for broadcast channels: the sender transmits independent quantum information to each of the receivers using entanglement he already shares with each of them. Like the original father protocol, it can easily be transformed into a protocol for entanglement-assisted transmission of classical information via superdense coding or into a protocol for unassisted transmission of qubits by using part of the transmission capacity to send the needed entanglement. Somewhat peculiarly, in this last case, the achievable rate region we obtain does not have an independent constraint on the sum-rate (unlike the third inequality of (\ref{eqn:marton})); instead, the two rates are separately upper-bounded by the coherent information. This might be seen as further evidence that the coherent information does not fully characterize the quantum capacity, or it might be due to the fact that the additional constraint we expect comes from the fact that we are looking at coherent information defined on a state of a slightly different form.

The paper is structured as follows. After introducing our notation and giving some background on quantum information in section \ref{sec:background}, as well as a quick review of the FQSW protocol in section \ref{sec:fqsw}, we present a high-level overview of the protocol in section \ref{sec:overview}. We then state and prove a one-shot version of the protocol in section \ref{sec:oneshot}, and then move on to the i.i.d. version of the protocol in section \ref{sec:iid}. Section \ref{sec:discussion} is devoted to a discussion of our results and outstanding issues.

\section{Background and notation}\label{sec:background}
Quantum subsystems will be labelled by capital letters $A$, $B$, etc; and their associated Hilbert spaces will be denoted by $\mathcal{H}_A$, $\mathcal{H}_B$, etc. When referring to a tensor product of $n$ isomorphic copies of a system $A$, we will write $A^n$. When necessary, we will use superscripts to indicate which subsystems a pure or mixed state is defined on; for instance, $\ket{\psi}^{AB} \in \mathcal{H}_{AB}$. We will abbreviate $\dim \mathcal{H}_A$ by $|A|$.

We will denote the partial trace by removing the corresponding system from the superscript; for instance, $\tr_A\left[ \rho^{AB} \right] = \rho^B$. Given a pure state $\ket{\psi}$, we will abbreviate its associated density matrix $\ketbra{\psi}$ by $\psi$.

Quantum operations will also be written using superscripts to denote the input and output systems; for example, $U^{A' \rightarrow B}$ is an operator which takes the quantum subsystem $A'$ as input and yields output on subsystem $B$. Generally, (partial) isometries will be written as $U$, $V$, and so forth, whereas quantum channels (also known as superoperators, or, more specifically, completely positive trace-preserving maps) will be written using calligraphic letters, such as $\mathcal{N}^{A' \rightarrow B}$. A quantum broadcast channel is a quantum channel with one input subsystem and two or more output subsystems.

Note that a quantum channel can always be extended to an isometry by adding another output subsystem which represents the environment of the channel (see, for example, \cite{holevo-stats}). This isometric extension implements exactly the same operation as the original channel if we trace out the environment subsystem. The isometric extension of $\mathcal{N}^{A' \rightarrow B}$ will be denoted by $U_{\mathcal{N}}^{A' \rightarrow BE}$, where $E$ is the environment. Note here that $U_{\mathcal{N}}$ does not act on density operators but on the Hilbert space $\mathcal{H}_{A'}$.

We denote conjugation of $B$ by $A$ using the symbol $\cdot$ in the form $A \cdot B := ABA^{\dagger}$. This will allow us to avoid writing symbols twice when applying several operators to a quantum state.

We will also denote a ``standard'' entangled pair between subsystems $S$ and $S'$ of equal size as $\ket{\Phi}^{SS'} = \tfrac{1}{\sqrt{|S|}}\sum_{i=0}^{|S|} \ket{ii}^{SS'}$, where the $\ket{i}^S$ and $\ket{i}^{S'}$ are some standard orthonormal bases on $S$ and $S'$.

We will often use the \emph{trace norm} of a Hermitian matrix $M$, defined to be $\left\| M \right\|_1 := \tr|M|$. It is particularly useful because it induces a statistically important metric on the space of quantum states; we call the quantity $\left\| \rho - \sigma \right\|_1$ the \emph{trace distance} between $\rho$ and $\sigma$. We will also use $\| M \|_0$ to refer to the rank of a matrix $M$.

The von Neumann entropy of a density operator $\rho^A$ will be denoted $H(\rho^A) = H(A)_\rho$. The quantum mutual information of $\rho^{AB}$ is the function $I(A;B)_\rho = H(A)_\rho + H(B)_\rho - H(AB)_{\rho}$ while the coherent information is the function $I(A\rangle B)_\rho = H(B)_\rho - H(AB)_\rho$.

Finally, we will say that two families of states $\psi$ and $\varphi$ parametrized by their size $n$ are asymptotically equal (denoted $\psi \aeq \varphi$) if $\left\| \psi - \varphi \right\|_1$ vanishes as $n \rightarrow \infty$. See Appendix \ref{sec:asymptotic} for a formal definition.

\subsection{Achievable rates and the capacity region}
\begin{figure}
    \centering
    \scalebox{0.6}{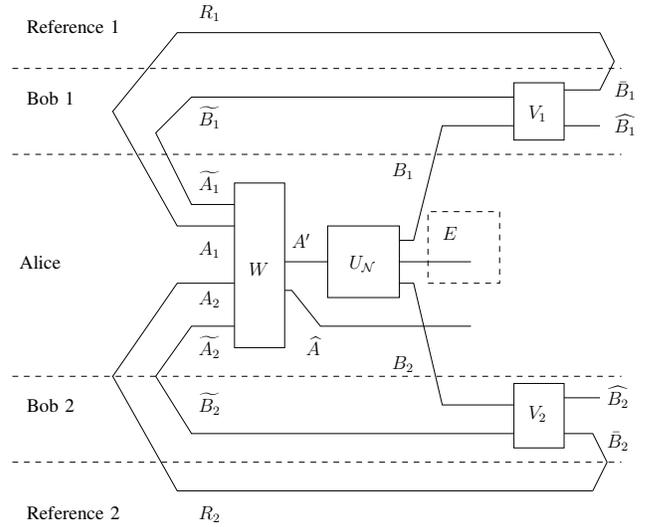}
    \caption{Diagram illustrating a generic protocol for a quantum broadcast channel. Each line represents a quantum system, boxes represent isometries, and the horizontal axis represents the passage of time. Lines joined together at either end of the diagram represent maximally entangled pairs.}
    \label{fig:achievable}
\end{figure}

Here we define what we mean by \emph{achievable rates} and the \emph{capacity region} of a quantum broadcast channel $\mathcal{N}^{A' \rightarrow B_1 B_2}$ for entanglement-assisted transmission. We define a \emph{$(Q_1, Q_2, n, \varepsilon)$-code} to consist of an encoding isometry $W^{A_1 \widetilde{A_1} A_2 \widetilde{A_2} \rightarrow \widehat{A} A'^{n}}$ and two decoding isometries $V_1^{B_1^{n} \widetilde{B_1} \rightarrow \bar{B}_1 \widehat{B_1}}$ and $V_2^{B_2^{n} \widetilde{B_2} \rightarrow \bar{B}_2 \widehat{B_2}}$ such that
\begin{equation*}
    \left\| \left( (V_2 V_1 U_{\mathcal{N}}^{\otimes n} W) \cdot \varphi \right) - \widehat{\psi}^{\widehat{B_1}\widehat{B_2}E\widehat{A}} \otimes \Phi^{R_1 \bar{B}_1} \otimes \Phi^{R_2 \bar{B}_2} \right\|_1 \leqslant \varepsilon
\end{equation*}
where $\ket{\varphi} = \ket{\Phi}^{R_1 A_1} \otimes \ket{\Phi}^{\widetilde{A_1} \widetilde{B_1}} \otimes \ket{\Phi}^{R_2 A_2} \otimes \ket{\Phi}^{\widetilde{A_2} \widetilde{B_2}}$ and $\widehat{\psi}^{\widehat{B_1}\widehat{B_2}E\widehat{A}}$ is a pure state, and where $\log|A_1| = Q_1$ and $\log|A_2| = Q_2$. $A_1$ and $A_2$ represent the systems that Alice wants to send to Bob 1 and Bob 2 respectively, and $\widetilde{A_1}\widetilde{B_1}$ and $\widetilde{A_2}\widetilde{B_2}$ are the EPR pairs Alice shares with the two receivers. See Figure \ref{fig:achievable} for a graphical illustration. Note that in practice, the encoding and decoding operations can be any completely positive, trace-preserving maps. We choose to implement these maps using isometries because this will prove much more convenient below.

A rate point $(Q_1,Q_2)$ is \emph{achievable} if there exists a sequence of $(Q_1, Q_2, n, \varepsilon_n)$-codes such that $\varepsilon_n \rightarrow 0$ as $n \rightarrow \infty$. The \emph{capacity region} of the channel $\mathcal{N}$ is the closure of the union of all achievable rate points.

Note that the protocols that we will be considering below also return some entanglement back to Alice and Bob 1 in addition to possibly Alice and Bob 2. This is of no consequence to the capacity region when we consider pre-shared entanglement to be free (as we have done above), but the reader can easily adapt the definition to accommodate this type of protocol.

The unassisted quantum capacity region for $\mathcal{N}$ is defined in the same way, except that the protocol begins without any entanglement between Alice and Bob 1 or Alice and Bob 2. Formally, the definitions are identical except that in the unassisted case,  the systems $\widetilde{A_1}$, $\widetilde{B_1}$, $\widetilde{A_2}$ and $\widetilde{B_2}$ are 1-dimensional or, equivalently, non-existent. This is where the amount of entanglement consumed by the protocols becomes crucial: the reduction from the entanglement-assisted case to the unassisted case involves using part of the transmission rate to generate the needed entanglement; therefore, every preshared ebit employed removes one qubit from the unassisted transmission rate.

\section{The FQSW protocol}\label{sec:fqsw}
Before presenting our protocol, we first give a quick overview of the fully quantum Slepian-Wolf protocol \cite{FQSW}. Suppose Alice and Bob hold a mixed state $\rho^{AB}$. We introduce a reference system $R$ to purify the state; the resulting state is $\ket{\psi}^{ABR}$. Alice would like to transfer her state to Bob by sending him as few qubits as possible. The FQSW theorem states that Alice can do this by first applying a unitary transformation to her entire share of the state (a random unitary selected according to the Haar measure will do with high probability), splitting her share into two subsystems $\bar{A}$ and $\widehat{A}$, and then sending $\widehat{A}$ to Bob.

Note that this scheme works provided that the subsystems $\bar{A}$ and $R$ are in a product state after applying the random unitary: since Bob holds the purifying system of $\bar{A}R$, there exists a local unitary that Bob can apply to turn his purifying system into separate purifying systems of the two subsystems. The purifying system of $R$ is exactly the original state that Alice wanted to send to Bob, and $\bar{A}$ together with its purifying system is an EPR pair shared by Alice and Bob. This last feature is an added bonus of the protocol: Alice and Bob get some free entanglement at the end.

It is possible to calculate how close $\bar{A}$ and $R$ are to being in a product state. The result of the calculation is the following (see \cite{FQSW} for details):
\begin{equation}
    \int_{\mathbb{U}(A)} \left\| \rho^{\bar{A} R}(U) - \frac{\ident^{\bar{A}}}{|\bar{A}|} \otimes \psi^R \right\|_1^2 dU \leqslant \frac{|A|\| \psi^R \|_0}{|\widehat{A}|^2} \tr\left[ \left( \psi^{AR} \right)^2 \right]
    \label{eqn:fqsw}
\end{equation}
where $\rho_{\bar{A}R}(U) = \tr_{\widehat{A}}[U \cdot \psi^{AR}]$. Since the inequality holds for the average over choices of $U$, there must exist at least one $U$ that satisfies it.

A special case of interest is when the initial state is an i.i.d. state of the form $(\ket{\psi}^{ABR})^{\otimes n}$. In this case, it can be shown that as long as $\log |\widehat{A}| \geqslant n[\frac{1}{2}I(A;R) + \delta]$ for some $\delta > 0$, it will be true that
\begin{equation}\label{eqn:asymptotic-fqsw}
\varphi^{\bar{A}R^{n}} \aeq \frac{\ident^{\bar{A}}}{|\bar{A}|} \otimes \varphi^{R^{n}}
\end{equation}
where $\varphi^{\bar{A}\widehat{A}B^{n}R^{n}}$ is the result of applying the random unitary to $\Pi_A \cdot (\psi^{ABR})^{\otimes n}$, where $\Pi_A$ is the projector onto the typical subspace of the $A$ subsystem, as defined in Appendix \ref{sec:typical}.

\section{Overview of the protocol}\label{sec:overview}
\begin{figure}
    \centering
    \scalebox{0.6}{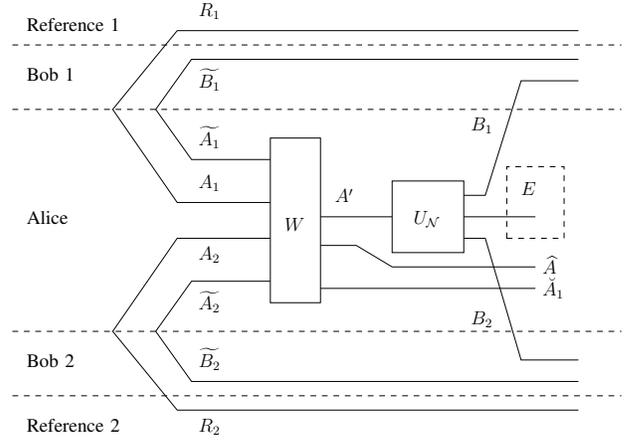}
    \caption{Diagram illustrating the one-shot version of the protocol.}
    \label{fig:bigbroadcast}
\end{figure}

Returning now to the broadcast setting, let's suppose Alice would like to send the maximally mixed system $A_1$ (which is purified by $R_1$) to Bob 1, and $A_2$ to Bob 2 using $n$ instances of the quantum broadcast channel $\mathcal{N}^{A' \rightarrow B_1 B_2}$. In addition, she has shared EPR pairs with both of them, represented by systems $\widetilde{A_1} \widetilde{B_1}$ and $\widetilde{A_2} \widetilde{B_2}$. At the end of the protocol, she will also retrieve some EPR pairs with Bob 1. Without loss of generality and to simplify the notation, we do not consider retrieving entanglement with Bob 2, since such a protocol could be simulated by timesharing with the same protocol with Bob 1 and Bob 2 reversed. We represent the channel by its isometric extension $U_{\mathcal{N}}^{A' \rightarrow B_1 B_2 E}$. Alice encodes her information using the encoding isometry $W^{A_1 \widetilde{A_1} A_2 \widetilde{A_2} \rightarrow A' \breve{A}_1 \widehat{A}}$; $A'$ is then transmitted through the channel, and $\widehat{A}$ is discarded (discarding a subsystem will turn out to be useful when discussing the i.i.d. case), and $\breve{A}_1$ eventually becomes Alice's share of the EPR pairs retrieved with Bob 1. Thus, after using the channel, the state of the system is $\ket{\psi} = U_{\mathcal{N}}^{\otimes n} W \ket{\varphi}$, where $\ket{\varphi} = \ket{\Phi}^{R_1 A_1} \otimes \ket{\Phi}^{\widetilde{A_1} \widetilde{B_1}} \otimes \ket{\Phi}^{R_2 A_2} \otimes \ket{\Phi}^{\widetilde{A_2} \widetilde{B_2}}$. See Figure \ref{fig:bigbroadcast} for a diagram illustrating this.

In order for Bob 1 to be able to decode, we have to make sure that $R_1 \breve{A}_1$ is in a product state with everything else that Bob 1 doesn't have access to, namely $R_2 B_2 \widetilde{B_2} E \widehat{A}$. Likewise, $R_2$ must be in a product state with $R_1 \breve{A}_1 B_1 \widetilde{B_1} E \widehat{A}$. This is accomplished by applying an FQSW random unitary on $R_1 \breve{A}_1 \widetilde{B_1}$ and another on $R_2 \widetilde{B_2}$, where $R_1 \breve{A}_1$ and $R_2$ each play the role of the system that stays behind. (Note that the choice of random unitary is made by all participants prior to initiating the protocol so does not require them to share any random bits.) Of course, it is impossible to apply these unitaries directly, since no one has access to $R_1$ and $R_2$, but we can note that $R_1 \widetilde{B_1} R_2 \widetilde{B_2}$ remains almost maximally mixed both before and after applying the FQSW unitaries, and that before using the channel, Alice has access to the system that purifies these subsystems. Hence, by Uhlmann's theorem, there exists a unitary that Alice can apply to her subsystems to achieve the same effect. The next section will argue this in more detail.

\section{One-shot version}\label{sec:oneshot}
We first prove a generic ``one-shot'' version of our theorem which works for general states and channels; we will then use it to derive an achievable rate region for the case of many independent uses of the channel.
\begin{thm}\label{thm:oneshot}
	For every encoding isometry $W^{A_1 \widetilde{A_1}A_2 \widetilde{A_2} \rightarrow A' \breve{A}_1 \widehat{A}}$, there exist an isometry $U^{A_1 \widetilde{A_1} A_2 \widetilde{A_2} \rightarrow A' \breve{A}_1 \widehat{A}}$ and decoding isometries $V_1^{B_1 \widetilde{B_1} \rightarrow \bar{B}_1 \widehat{B_1} \breve{B}_1}$ and $V_2^{B_2 \widetilde{B_2} \rightarrow \bar{B}_2 \widehat{B_2}}$ such that for all states $\psi_1$ and $\psi_2$ defined on $R_1 R_2 \widetilde{B_1} \widetilde{B_2} \breve{A}_1 B_1 B_2 \widehat{A} E$,
\begin{multline}
\label{eqn:oneshot-final}
\left\| \left( (V_2 V_1 U_{\mathcal{N}} U) \cdot \varphi \right) - \widehat{\psi}^{\widehat{B_1}\widehat{B_2}E\widehat{A}} \otimes \Phi^{\breve{A}_1 \breve{B}_1} \otimes \Phi^{R_1 \bar{B}_1} \otimes \Phi^{R_2 \bar{B}_2} \right\|_1\\
\leqslant 4\left\{ \frac{|R_1||\widetilde{B_1}||\breve{A}_1|\|\psi_1^{R_2\widetilde{B_2}B_2 E\widehat{A}}\|_0}{|\widetilde{B_1}|^2} \tr[(\psi_1^{R_1 \widetilde{B_1} \breve{A}_1 R_2 \widetilde{B_2} B_2 E \widehat{A}})^2] \right\}^{\frac{1}{4}}\\
+ 2\left\{ \frac{|R_2||\widetilde{B_2}|\|\psi_2^{R_1\widetilde{B_1}\breve{A}_1 B_1 E\widehat{A}}\|_0}{|\widetilde{B_2}|^2} \tr[(\psi_2^{R_2 \widetilde{B_2} R_1 \widetilde{B_1} \breve{A}_1 B_1 E \widehat{A}})^2] \right\}^{\frac{1}{4}}\\
+ 2\| \psi - \psi_1 \|_1 + \| \psi - \psi_2 \|_1
\end{multline}
where $\ket{\varphi} = \ket{\Phi}^{R_1 A_1} \otimes \ket{\Phi}^{\widetilde{A_1} \widetilde{B_1}} \otimes \ket{\Phi}^{R_2 A_2} \otimes \ket{\Phi}^{\widetilde{A_2} \widetilde{B_2}}$, $\ket{\psi} = U_{\mathcal{N}} W \ket{\varphi}$, and $\widehat{\psi}^{\widehat{B_1}\widehat{B_2}E\widehat{A}}$ is a pure state uniquely determined by the protocol.
\end{thm}
Here, one should think of $W$ merely as an arbitrary way to map inputs to the protocol (qubits we want to transmit and preshared entanglement) into inputs to the channel; it need not have any actual error-correction capability. Our theorem uses it to create $U$, a ``new and improved'' version of $W$ which does have error-correction capabilities, along with its associated decoders $V_1$ and $V_2$.
\begin{proof}
Applying formula (\ref{eqn:fqsw}) twice, once on $\psi_1$ with a random unitary over $R_1 \widetilde{B_1} \breve{A}_1$ and once on $\psi_2$ with a random unitary over $R_2 \widetilde{B_2}$, yields:
\begin{multline}
	\hspace{-5mm}	\int \left\| \sigma_1^{R_1 \breve{A}_1 R_2 \widetilde{B_2} B_2 E\widehat{A}}(U) - \frac{\ident^{R_1 \breve{A}_1}}{|R_1||\breve{A}_1|} \otimes \psi_1^{R_2 \widetilde{B_2} B_2 E\widehat{A}} \right\|^2_1 dU\\
	\leqslant \frac{|R_1||\breve{A}_1||\widetilde{B_1}|\|\psi_1^{R_2\widetilde{B_2}B_2E\widehat{A}}\|_0}{|\widetilde{B_1}|^2}\tr\left[(\psi_1^{R_1 \widetilde{B_1} \breve{A}_1 R_2 \widetilde{B_2} B_2 E \widehat{A}})^2\right]
\end{multline}
where the integral is taken oven random unitaries on $R_1 \widetilde{B_1} \breve{A}_1$ and $\sigma_1^{R_1 \breve{A}_1 R_2 \widetilde{B_2} B_2 E \widehat{A}}(U) = \tr_{\widetilde{B_1}}[U \cdot \psi_1^{R_1 \widetilde{B_1} \breve{A}_1 R_2 \widetilde{B_2} B_2 E \widehat{A}}]$, and
\begin{multline}
	\hspace{-0.6cm}    \int \left\| \sigma_2^{R_2 R_1 \widetilde{B_1} \breve{A}_1 B_1 E \widehat{A}}(U) - \frac{\ident_{R_2}}{|R_2|} \otimes \psi_2^{R_1 \widetilde{B_1} \breve{A}_1 B_1 E \widehat{A}} \right\|^2_1 dU\\
	\leqslant \frac{|R_2||\widetilde{B_2}| \|\psi_2^{R_1\widetilde{B_1}\breve{A}_1 B_1E\widehat{A}}\|_0}{|\widetilde{B_2}|^2} \tr[(\psi_2^{R_2 \widetilde{B_2} R_1 \widetilde{B_1} \breve{A}_1 B_1 E \widehat{A}})^2].
\end{multline}
where the integral is taken oven random unitaries on $R_2 \widetilde{B_2}$ and $\sigma_2^{R_2 R_1 \widetilde{B_1} \breve{A}_1 B_1 E \widehat{A}}(U) = \tr_{\widetilde{B_2}}[U \cdot \psi_2^{R_2 \widetilde{B_2} R_1 \widetilde{B_1} \breve{A}_1 B_1 E \widehat{A}}]$.

This means that there exist unitaries $U_1^{R_1 \breve{A}_1 \widetilde{B_1}}$ and $U_2^{R_2 \widetilde{B_2}}$ that satisfy the above inequalities. Now, since $W \cdot \varphi$ is asymptotically equal to the maximally mixed state over $R_1 \widetilde{B_1} R_2 \widetilde{B_2}$ both before and after applying $U_1$ and $U_2$ to it, by Uhlmann's theorem (see Appendix \ref{sec:uhlmann}) there must exist some unitary transformation $U'$ on $A' \breve{A}_1 \widehat{A}$ which achieves the same effect; and since Alice has these three subsystems in her possession before using the channel, she can perform this transformation. Thus, let $U = U' W$.

Now, using Uhlmann's theorem once again, we get that there exist decoding unitaries $V_1^{B_1 \widetilde{B_1} \rightarrow \bar{B}_1 \breve{B}_1 \widehat{B_1}}$ and $V_2^{B_2 \widetilde{B_2} \rightarrow \bar{B}_2 \widehat{B_2}}$ such that
\begin{multline}
\label{eqn:oneshot-one}
\hspace{-5mm}\left\| \left( (V_2 V_1 U_{\mathcal{N}} U) \cdot \varphi \right) - \widehat{\psi}_1^{R_2 \bar{B}_2 \widehat{B_1} \widehat{B_2} E\widehat{A}} \otimes \Phi^{\breve{A}_1 \breve{B}_1} \otimes \Phi^{R_1 \bar{B}_1} \right\|_1\\
\hspace{-3mm}\leqslant 2\left\{ \frac{|R_1\breve{A}_1\widetilde{B_1}|\|\psi_1^{R_2\widetilde{B_2}B_2 E\widehat{A}}\|_0}{|\widetilde{B_1}|^2} \tr[(\psi_1^{R_1 \widetilde{B_1} \breve{A}_1 R_2 \widetilde{B_2} B_2 E \widehat{A}})^2] \right\}^{\frac{1}{4}}
\end{multline}

and

\begin{multline}
\label{eqn:oneshot-two}
\left\| \left( (V_2 V_1 U_{\mathcal{N}}U) \cdot \varphi \right) - \widehat{\psi}_2^{R_1 \breve{A}_1 \bar{B}_1 \widehat{B_1} \widehat{B_2} E \widehat{A}} \otimes \Phi^{R_2 \bar{B}_2} \right\|_1\\
\leqslant 2\left\{ \frac{|R_2\widetilde{B_2}|\|\psi_2^{R_1\widetilde{B_1}\breve{A}_1 B_1 E\widehat{A}}\|_0}{|\widetilde{B_2}|^2} \tr[(\psi_2^{R_2 \widetilde{B_2} R_1 \widetilde{B_1} \breve{A}_1 B_1 E \widehat{A}})^2] \right\}^{\frac{1}{4}}
\end{multline}
where $\widehat{\psi}_1$ and $\widehat{\psi}_2$ are some pure states determined by the theorem.
To finish, we apply the following lemma and the triangle inequality and obtain equation (\ref{eqn:oneshot-final}).
\begin{lemma}
	If we have density operators $\rho^{ABC}, \sigma^A, \omega^{BC}, \tau^{AB}, \eta^C$ such that
\begin{align*}
\left\| \rho^{ABC} - \sigma^A \otimes \omega^{BC} \right\|_1 &\leqslant \varepsilon_1\\
\left\| \rho^{ABC} - \tau^{AB} \otimes \eta^C \right\|_1 &\leqslant \varepsilon_2
\end{align*}
then $\left\| \rho^{ABC} - \sigma^A \otimes \tau^B \otimes \eta^C \right\|_1 \leqslant 2\varepsilon_1 + \varepsilon_2$.
\end{lemma}
\vskip 2mm
\emph{Proof of lemma:}
\begin{multline*}
\left\| \rho^{ABC} - \sigma^A \otimes \tau^B \otimes \eta^C \right\|_1\\
\begin{split}
&\leqslant \left\| \rho^{ABC} - \sigma^A \otimes \omega^{BC} \right\|_1\\
&\quad + \left\| \sigma^A \otimes \omega^{BC} - \sigma^A \otimes \tau^B \otimes \eta^C \right\|_1\\
&= \varepsilon_1 + \left\| \omega^{BC} - \tau^B \otimes \eta^C \right\|_1\\
&\leqslant \varepsilon_1 + \left\| \omega^{BC} - \rho^{BC} \right\|_1 + \left\| \rho^{BC} - \tau^B \otimes \eta^C \right\|_1\\
&\leqslant 2\varepsilon_1 + \varepsilon_2
\end{split}
\end{multline*}
where the first two inequalities are applications of the triangle inequality, and the equality is due to the fact that $\| A \|_1 = \| \sigma \otimes A \|_1$ for any operator $A$ and density matrix $\sigma$. 
\end{proof}

\section{i.i.d version}\label{sec:iid}
\begin{thm}\label{thm:iid}
	Let $\mathcal{N}^{A' \rightarrow B_1 B_2}$ be a quantum broadcast channel. Then, for every pure state $\ket{\phi}^{A_1 A_2 A' D}$, the following rate region is achievable for entanglement-assisted transmission:
\begin{equation}
\label{eqn:iidregion}
\begin{split}
0 \leqslant Q_1 &\leqslant \frac{1}{2} I(A_1; B_1)_\psi\\
0 \leqslant Q_2 &\leqslant \frac{1}{2} I(A_2; B_2)_\psi\\
Q_1 + Q_2 &\leqslant \frac{1}{2} \left[ I(A_1;B_1)_\psi + I(A_2;B_2)_\psi - I(A_1;A_2)_\psi \right].
\end{split}
\end{equation}
$Q_1$ is the rate at which Alice sends qubits to Bob 1, and likewise for $Q_2$ for Bob 2, and $\ket{\psi}^{A_1 A_2 B_1 B_2 D E} = U_{\mathcal{N}}^{A' \rightarrow B_1 B_2 E}\ket{\phi}^{A_1 A_2 A' D}$.
\end{thm}

Note that including the $D$ subsystem is equivalent to allowing $\phi^{A_1 A_2 A'}$ to be a mixed state; we find this formulation more convenient for our purposes.

\begin{proof}
	To get this rate region, we must apply the one-shot theorem to an i.i.d. state. The main challenge is that for an arbitrary i.i.d. state of the form $(\ket{\psi}^{A_1 A_2 B_1 B_2 D E})^{\otimes n} = U_{\mathcal{N}}^{\otimes n} (\ket{\phi}^{A_1 A_2 A' D})^{\otimes n}$, the $A_1^{n}$ and $A_2^{n}$ subsystems can be correlated, and to apply the one-shot theorem, it is crucial that $A_1^{n}$ and $A_2^{n}$ be maximally mixed and decoupled in order to play the roles of $R_1 \widetilde{B_1}$ and $R_2 \widetilde{B_2}$ respectively. (We use the term \emph{decoupled} to indicate that the density operator of a composite quantum system is the product of the reduced density operators of its component systems. The analogous notion in probability theory is independence.)

    We can remedy this situation by using the FQSW protocol to decouple $A_1^{n}$ and $A_2^{n}$. Whether we apply it to $A_1^{n}$ or to $A_2^{n}$, it will require us to remove $n[\frac{1}{2} I(A_1; A_2) + \delta]$ qubits, where $\delta > 0$ can be arbitrarily small. (Note that here, and throughout this proof, the mutual information is taken with respect to $\ket{\psi}$ as defined in the statement of the theorem.) The removed qubits will play the role of $\breve{A}_1$ in the previous section and represent halves of EPR pairs generated with Bob 1 at the end of the protocol. Suppose without loss of generality that we apply FQSW to $A_1^{n}$ only. (This will correspond to one of the corner points of the region and therefore, by time-sharing, the entire region will be achievable.) Let $W_1^{A_1^{n} \rightarrow \bar{A}_1 \breve{A}_1 \widehat{A_1}}$ be a Schumacher compression isometry (meaning an operator that separates the typical and non-typical subspaces into distinct subsystems) composed with this FQSW unitary where $\bar{A}_1$ plays the role of the system that stays behind in FQSW, $\breve{A}_1$ is the system that is discarded in the FQSW step, and $\widehat{A_1}$ is the system that is discarded in the compression step.

    At the end of this process, by equation (\ref{eqn:asymptotic-fqsw}), the $\bar{A}_1$ subsystem of $W_1 \cdot \psi^{\otimes n}$ is asymptotically equal to the maximally mixed state. To get $A_2^{n}$ to also be maximally mixed, we can apply another FQSW unitary to it (after Schumacher compressing it), and discard $n\delta$ qubits from it (where $\delta$ can be arbitrarily small); this also leaves $\bar{A}_2$ asymptotically equal to the maximally mixed state. Let $W_2^{A_2^{n} \rightarrow \bar{A}_2 \widehat{A_2}}$ be a Schumacher compression unitary followed by this second FQSW unitary as with $W_1$, and let $\ket{\xi}^{\bar{A}_1\bar{A}_2 \breve{A}_1 \widehat{A_1} \widehat{A_2} {A'}^n D^{n}}$ be
\begin{equation*}
	W_2^{A_2^{n} \rightarrow \bar{A}_2 \widehat{A_2}} W_1^{A_1^{n} \rightarrow \bar{A}_1 \breve{A}_1 \widehat{A_1}} (\ket{\phi}^{A_1 A_2 A' D})^{\otimes n}.
\end{equation*}
Applying equation (\ref{eqn:asymptotic-fqsw}) to $W_1$ and $W_2$, we obtain that
\begin{align}
    \xi^{\bar{A}_1\bar{A}_2\widehat{A_2}} &\aeq \frac{\ident^{\bar{A}_1}}{|\bar{A}_1|} \otimes \xi^{\bar{A}_2\widehat{A_2}}\\
    \xi^{\bar{A}_2} &\aeq \frac{\ident^{\bar{A}_2}}{|\bar{A}_2|}
\end{align}
Hence, we have that $\xi^{\bar{A}_1\bar{A}_2} \aeq \frac{\ident^{\bar{A}_1\bar{A}_2}}{|\bar{A}_1||\bar{A}_2|}$, confirming that $\bar{A}_1\bar{A}_2$ is indeed maximally mixed.

Now, let $\ket{\varphi} = \ket{\Phi}^{R_1 A_1} \otimes \ket{\Phi}^{\widetilde{A_1} \widetilde{B_1}} \otimes \ket{\Phi}^{R_2 A_2} \otimes \ket{\Phi}^{\widetilde{A_2} \widetilde{B_2}}$, where we identify $R_1 \widetilde{B_1}$ with $\bar{A}_1$ and $R_2 \widetilde{B_2}$ with $\bar{A}_2$. Since the state on $\bar{A}_1\bar{A}_2$ is asymptotically equal to the maximally mixed state in both $\ket{\xi}$ and $\ket{\varphi}$, by Uhlmann's theorem (see Appendix \ref{sec:uhlmann}) there exists an isometry $W^{A_1 \widetilde{A_1} A_2 \widetilde{A_2} \rightarrow \widehat{A_1} \widehat{A_2} \breve{A}_1 A'^n D^{n}}$ such that $\ket{\xi_U} := W \ket{\varphi}$ is asymptotically equal to $\ket{\xi}$. Note that we can use Theorem \ref{thm:oneshot} directly on $\ket{\varphi}$ and the encoding unitary $W$. This means that there exist isometries $U^{A_1 \widetilde{A_1} A_2 \widetilde{A_2} \rightarrow \widehat{A_1} \widehat{A_2} \breve{A}_1 A'^n D^{n}}$, $V_1^{B_1 \widetilde{B_1} \rightarrow \bar{B}_1 \breve{B}_1 \widehat{B_1}}$, and $V_2^{B_2 \widetilde{B_2} \rightarrow \bar{B}_2 \widehat{B_2}}$ such that equation (\ref{eqn:oneshot-final}) is satisfied if we identify $\widehat{A_1} \widehat{A_2} D^{n}$ with the subsystem $\widehat{A}$ in Theorem \ref{thm:oneshot}.

Now, define $\Pi_F$ to be the projector onto the $\varepsilon(n)$-typical subspace of an arbitrary subsystem $F^{n}$ (see Appendix \ref{sec:typical}). Let $\ket{\xi_1}$ be
\begin{equation*}
    W_2 W_1 {U_{\mathcal{N}}^{\otimes n}}^{\dagger}  \Pi_{A_2 B_2 DE} \Pi_{A_1} \Pi_{A_1 A_2 B_2 DE} \ket{\psi}^{\otimes n}
\end{equation*}
and $\ket{\xi_2}$ be
\begin{equation*}
    W_2 W_1 {U_{\mathcal{N}}^{\otimes n}}^{\dagger}  \Pi_{A_1 B_1 DE} \Pi_{A_2} \Pi_{A_2 A_1 B_1 DE} \ket{\psi}^{\otimes n}.
\end{equation*}
Since the only differences between $\ket{\xi}$, $\ket{\xi_1}$ and $\ket{\xi_2}$ are the presence of different typical projectors, it is possible (see Appendix \ref{sec:typical}, or Appendix A in \cite{FQSW}) to choose $\varepsilon(n)$ such that $\lim_{n\rightarrow\infty} \varepsilon(n) = 0$ and such that the three states are asymptotically equal. (Note that the argument relies on the transitivity of asymptotic equality.) We will therefore select $\varepsilon(n)$ such that $\xi \aeq \xi_U \aeq \xi_1 \aeq \xi_2$.

We will now evaluate the right-hand side of (\ref{eqn:oneshot-final}) using $\xi_{U,\mathcal{N}} = U_{\mathcal{N}}^{\otimes n} \cdot \xi_U$ as $\psi$, $\xi_{1,\mathcal{N}} = U_{\mathcal{N}}^{\otimes n} \cdot \xi_1$ as $\psi_1$, and $\xi_{2,\mathcal{N}} = U_{\mathcal{N}}^{\otimes n} \cdot \xi_2$ as $\psi_2$. Note that $\bar{A}_1$ will be split into $R_1$ and $\widetilde{B_1}$ and likewise for $\bar{A}_2$. From basic properties of typical subspaces (see Appendix \ref{sec:typical}), for sufficiently large $n$ we have:
\begin{equation}
	|R_1||\breve{A}_1||\widetilde{B_1}| \leqslant 2^{n[H(A_1) + \delta]}\\
\end{equation}
since $R_1$, $\breve{A}_1$ and $\widetilde{B_1}$ taken together form the typical subspace of $A_1$, which is of size $2^{n[H(A_1)+\delta]}$. We also have
\begin{equation}
\begin{split}
	\left\|\xi_{1,\mathcal{N}}^{R_2 \widetilde{B_2} B_2^{n} D^{n}E^{n} \widehat{A_2}}\right\|_0 &= \left\|\Pi_{A_2 B_2 D E}\right\|_0\\
      &\leqslant 2^{n[H(A_2 B_2 DE) + \delta]}\\
\end{split}
\end{equation}
Finally, we have
\begin{multline}
	\tr\left[ \left( \xi_{1,\mathcal{N}}^{\bar{A}_1\bar{A}_2 \breve{A}_1 B_2^{n} D^{n}E^{n} \widehat{A_1}\widehat{A_2}}\right)^2 \right]\\
\begin{split}
	&= \tr\left[ \left( (W_1^{\dagger}W_2^{\dagger} U_{\mathcal{N}}^{\otimes n} \cdot \xi_{1,\mathcal{N}})^{A_1^{n} A_2^{n} B_2^{n} D^{n}E^{n}}\right)^2 \right]\\
	&= \tr\left[\left(\Pi_{A_2B_2DE}\Pi_{A_1}\Pi_{A_1A_2B_2DE}U_{\mathcal{N}}^{\otimes n} \cdot \phi^{\otimes n}\right)^2\right]\\
	&\leqslant \tr\left[\left(\Pi_{A_1A_2B_2DE}U_{\mathcal{N}}^{\otimes n} \cdot \phi^{\otimes n}\right)^2\right]\\
     &\leqslant 2^{-n[H(A_1 A_2 B_2 DE) - \delta]}
\end{split}
\end{multline}
where we used the definition of $\xi_{1,\mathcal{N}}$ in the second equation, and the first inequality is due to the fact that adding a projector can only decrease the trace. Therefore, the first term of equation (\ref{eqn:oneshot-final}) becomes
\begin{multline*}
	4\bigg\{ \frac{|R_1||\breve{A}_1||\widetilde{B_1}|\|(\xi_{1,\mathcal{N}})^{R_2\widetilde{B_2}B_2^{n} D^{n}E^{n}\widehat{A}}\|_0}{|\widetilde{B_1}|^2}\\
	\tr\left[  \left( (\xi_{1,\mathcal{N}})^{R_1 \widetilde{B_1} R_2 \widetilde{B_2} B_2^{n} \breve{A}_1 D^{n}E^{n} \widehat{A}} \right)^2  \right]\bigg\}^{\frac{1}{4}}\\
    \leqslant 4\left\{ \frac{2^{n[I(A_1;A_2 B_2 DE) + 3\delta]}}{|\widetilde{B_1}|^2} \right\}^{\frac{1}{4}}
\end{multline*}

Assuming $|\widetilde{B_1}| \geqslant 2^{n[I(A_1;A_2 B_2 DE)/2 + 2\delta]}$, we get
\begin{multline*}
	4\bigg\{ \frac{|R_1||\breve{A}_1||\widetilde{B_1}|\|(\xi_{1,\mathcal{N}})^{R_2\widetilde{B_2}B_2^{n} D^{n}E^{n}\widehat{A}}\|_0}{|\widetilde{B_1}|^2}\\
	\tr \left[ \left( (\xi_{1,\mathcal{N}})^{R_1 \widetilde{B_1} R_2 \widetilde{B_2} \breve{A}_1 B_2^{n} D^{n}E^{n} \widehat{A}}\right]^2 \right)  \bigg\}^{\frac{1}{4}}\\
    \leqslant 4\times 2^{-n\delta/4}
\end{multline*}

Likewise, we can evaluate the second term on the right-hand side of equation (\ref{eqn:oneshot-final}) and conclude that we need $|\widetilde{B_2}| \geqslant 2^{n[I(A_2; A_1 B_1 DE)/2 + 2\delta]}$ to make it vanish. The third and fourth terms then vanish due to the fact that $\ket{\xi_{1,\mathcal{N}}} \aeq \ket{\xi_{2,\mathcal{N}}} \aeq U_{\mathcal{N}}^{\otimes n} \ket{\xi_U}$. Hence, we get that
\begin{multline*}
	(V_2 V_1 U_{\mathcal{N}}^{\otimes n} W U_2 U_1) \cdot \varphi\\
	\aeq \widehat{\psi}^{\widehat{B_1}\widehat{B_2}D^{n}E^{n}\widehat{A}} \otimes \Phi^{R_1 \bar{B}_1} \otimes \Phi^{R_2 \bar{B}_2} \otimes \Phi^{\breve{A}_1 \breve{B}_1},
\end{multline*}
which means that the protocol works.

We can now easily verify that our conditions on $|\widetilde{B_1}|$ and $|\widetilde{B_2}|$ indeed correspond to the rates advertised in the statement of the theorem. First, we have
\begin{equation*}
\begin{split}
nQ_1 &= \log|R_1|\\
&=  \log|\bar{A}_1| - \log|\widetilde{B_1}|\\
&\leqslant n\left[ H(A_1) - \frac{1}{2}I(A_1;A_2) - \frac{1}{2}I(A_1;A_2 B_2 DE) - 3\delta \right]\\
&= \frac{1}{2}n\left[ I(A_1;B_1) - I(A_1;A_2) - 3\delta \right]
\end{split}
\end{equation*}
and
\begin{equation*}
\begin{split}
    nQ_2 = \log|R_2| &=  \log|\bar{A}_2| - \log|\widetilde{B_2}|\\
&\leqslant n\left[ H(A_2) - \frac{1}{2}I(A_2;A_1 B_1 DE) - 2\delta \right]\\
&= \frac{1}{2}n\left[ I(A_2;B_2) - 2\delta \right]
\end{split}
\end{equation*}
where $\delta$ vanishes as $n \rightarrow \infty$. We can, of course, exchange the roles of Bob 1 and Bob 2; combining this with time-sharing gives the asymptotic rates given in (\ref{eqn:iidregion}).
\end{proof}

We can also calculate how much entanglement is needed between Alice and the two Bobs; let $E_1$ be the rate at which EPR pairs between Alice and Bob 1 are used during the protocol, and define $E_2$ similarly for Bob 2. Since entanglement is created between Alice and Bob 1 at the end of the protocol, we take the difference between the rate consumed by the protocol and the rate at which entanglement is recreated at the end. We have
\begin{equation}
\begin{split}
	n E_1 &= \log |\widetilde{B_1}| - \log |\breve{A}_1|\\
	&\geqslant n \left[ \frac{1}{2} I(A_1; A_2 B_2 DE) - \frac{1}{2} I(A_1; A_2) + \delta \right]\\
	&= n \left[ \frac{1}{2} I(A_1;B_2 D E | A_2) + \delta \right]\\
    n E_2 &= \log |\widetilde{B_2}|\\
    &\geqslant n \left[ \frac{1}{2} I(A_2; A_1 B_1 DE) + \delta \right]
\end{split}
\end{equation}

\subsection{Unassisted transmission}
Note that a simple modification of this protocol allows us to transmit qubits without needing preshared entanglement. We can first let Alice establish initial entanglement with Bob 1 using the LSD Theorem \cite{lsd1, lsd2, lsd3} (ignoring Bob 2 during this phase of the protocol); likewise, she can establish initial entanglement with Bob 2. Then, they can use the entanglement-assisted protocol just shown for the rest of the transmission, using part of the rate to maintain their stock of entanglement, and using the surplus to transmit qubits. In other words, they voluntarily downgrade part of the transmission rate to entanglement generation. Since we only need to use this suboptimal protocol for the initial stage, the asymptotic rates will be unaffected. The asymptotic rates will be
\begin{equation*}
\begin{split}
    \bar{Q}_1 &= Q_1 - E_1\\
    &\leqslant \frac{1}{2} I(A_1; B_1) - \frac{1}{2} I(A_1; A_2) - \frac{1}{2} I(A_1; B_2 DE|A_2)\\
    &= I(A_1 \rangle B_1)\\
    \bar{Q}_2 &= Q_2 - E_2\\
    &\leqslant \frac{1}{2} I(A_2; B_2) - \frac{1}{2} I(A_2; A_1 B_1 DE)\\
    &= I(A_2 \rangle B_2)\\
\end{split}
\end{equation*}
yielding, via time-sharing, the following rate region:
\begin{equation*}
\begin{split}
    0 \leqslant \bar{Q}_1 &\leqslant I(A_1 \rangle B_1)\\
    0 \leqslant \bar{Q}_2 &\leqslant I(A_2 \rangle B_2)\\
\end{split}
\end{equation*}
It is remarkable that in the case of unassisted transmission, we do not get a ``penalty term'' on the sum rate; the two individual rates are constrained separately by an expression having exactly the same form as for transmission over point-to-point channels. We can see that this is due to the fact that the part of $A_1$ that is discarded in order to decouple it from $A_2$ is not lost: instead of contributing to the transmission rate, it is simply ``downgraded'' to entanglement generation and is therefore just as useful for regenerating the entanglement needed by the entanglement-assisted protocol. However, standard techniques for converting entanglement generation protocols into quantum transmission protocols (see for instance \cite{BKN98}) cannot be used profitably here, since this additional transmission rate would have to be used to regenerate the entanglement stock anyway. 

A detailed proof that this strategy succeeds without any initial investment of entanglement requires a slightly more careful analysis of the broadcast father protocol than we have done here. Specifically, it is straightforward to verify that the entanglement generated in the father can be produced such that it is within $O(2^{-n\alpha})$ in trace distance of the standard maximally entangled state, for some $\alpha > 0$. This ensures that the father protocol can be repeated a  number of times polynomial in $n$, re-using some of the output entanglement at each step, without causing significant degradation in the quality of the entanglement.

\subsection{Regularized converse}
The rate region given in Theorem \ref{thm:iid} is indeed the capacity of quantum broadcast channels provided we regularize over many uses of the channel. It is important to remember, however, that regions defined by very different formulas can nonetheless agree after regularization, so the following theorem should be understood to be only a very weak characterization of the capacity.
\begin{thm}\label{thm:reg-converse}
    The entanglement-assisted capacity region of a quantum broadcast channel $\mathcal{N}^{A' \rightarrow B_1 B_2}$ is the convex hull of the union of all rate points $(Q_1, Q_2)$ satisfying
\begin{equation}\label{eqn:converse-region}
\begin{split}
    0 \leqslant Q_1 &\leqslant \frac{1}{2n} I(A_1; B_1^{n})\\
    0 \leqslant Q_2 &\leqslant \frac{1}{2n} I(A_2; B_2^{n})\\
    Q_1 + Q_2 &\leqslant \frac{1}{2n} [I(A_1; B_1^{n}) + I(A_2; B_2^{n}) - I(A_1;A_2)]\\
\end{split}
\end{equation}
for some state of the form $\ket{\psi}^{A_1 A_2 B_1^{n} B_2^{n} D E^{n}} = {U^{\otimes n}_{\mathcal{N}}}\ket{\phi}^{A_1 A_2 {A'}^{n} D}$, where  $\ket{\phi}$ is a pure state.
\end{thm}
\begin{proof}
    It is immediate from Theorem \ref{thm:iid} that the region is achievable. We now prove the converse.

    Suppose that $(Q_1, Q_2)$ is an achievable rate pair. That means that there exists a sequence of $(Q_1, Q_2, n, \varepsilon_n)$ codes such that $\varepsilon_n \rightarrow 0$ as $n \rightarrow \infty$. Consider the code of block size $n$ in this sequence. Let $\ket{\varphi} = \ket{\Phi}^{R_1 A_1} \otimes \ket{\Phi}^{\widetilde{A_1} \widetilde{B_1}} \otimes \ket{\Phi}^{R_1 A_1} \otimes \ket{\Phi}^{\widetilde{A_1} \widetilde{B_1}}$ be the input state as in Theorem \ref{thm:oneshot}, $W^{A_1 A_2 \widetilde{A_1}\widetilde{A_2} \rightarrow A'^{n}D}$ be the encoding isometry, and let $\ket{\psi}^{R_1 R_2 B_1^{n} B_2^{n} \widetilde{B_1} \widetilde{B_2} E^{n} D^{n}} = U_{\mathcal{N}}^{\otimes n} W \ket{\varphi}$. As usual, we will evaluate entropic quantities with respect to $\ket{\psi}$.

    Given that Bob 1 must be able to recover a system which purifies $R_1$ from $B_1^{n}$ and $\widetilde{B_1}$, we have by Fannes' inequality \cite{fannes} that $I(R_1; B_1^{n} \widetilde{B_1}) \geqslant 2\log|R_1| - n\delta_n$, where $\delta_n \rightarrow 0$ as $n \rightarrow \infty$, and likewise for Bob 2. We also have
\begin{equation}
\begin{split}
    I(R_1; B_1^{n}\widetilde{B_1}) &= H(R_1) + H(B_1^{n} \widetilde{B_1}) - H(R_1 B_1^{n} \widetilde{B_1})\\
    &\leqslant H(R_1) + H(B_1^{n})\\
    &\quad + H(\widetilde{B_1}) - H(R_1 B_1^{n} \widetilde{B_1})\\
    &= H(R_1\widetilde{B_1}) + H(B_1^{n}) - H(R_1 B_1^{n} \widetilde{B_1})\\
    &= I(R_1 \widetilde{B_1};B_1^{n})
\end{split}
\end{equation}
where the second line follows from subadditivity, and the third line from the fact that $R_1$ and $\widetilde{B_1}$ are in a product state. Hence, $I(R_1 \widetilde{B_1}; B_1^{n}) \geqslant 2\log|R_1| - n\delta_n$ and likewise, $I(R_2 \widetilde{B_2}; B_2^{n}) \geqslant 2\log|R_2| - n\delta_n$. Now, if we identify $R_1 \widetilde{B_1}$ as $A_1$ and $R_2 \widetilde{B_2}$ as $A_2$, we see that
\begin{align}
    Q_1 &\leqslant \frac{1}{2n} I(A_1; B_1^{n}) + \delta_n\\
    Q_2 &\leqslant \frac{1}{2n} I(A_2; B_2^{n}) + \delta_n
\end{align}
where $\delta_n \rightarrow 0$ as $n \rightarrow \infty$. Since $I(A_1;A_2) = 0$, this rate point is clearly inside the region in equation (\ref{eqn:converse-region}), and it follows that this is indeed the capacity of the channel.
\end{proof}

An analogous theorem can easily be shown to hold for the unassisted capacity:
\begin{thm}\label{thm:reg-converse-unassisted}
    The unassisted capacity region of a quantum broadcast channel $\mathcal{N}^{A' \rightarrow B_1 B_2}$ is the convex hull of the union of all rate points $(Q_1, Q_2)$ satisfying
\begin{equation}\label{eqn:converse-region-unassisted}
\begin{split}
    0 \leqslant Q_1 &\leqslant \frac{1}{n} I(A_1 \rangle B_1^{n})\\
    0 \leqslant Q_2 &\leqslant \frac{1}{n} I(A_2 \rangle B_2^{n})\\
\end{split}
\end{equation}
for some state of the form $\ket{\psi}^{A_1 A_2 B_1^{n} B_2^{n} D E^{n}} = {U^{\otimes n}_{\mathcal{N}}}\ket{\phi}^{A_1 A_2 {A'}^{n} D}$, where  $\ket{\phi}$ is a pure state.
\end{thm}
While one might conjecture that Theorem \ref{thm:reg-converse} characterizes the entanglement-assisted capacity region of a broadcast channel even with the restriction $n=1$, the analogous conjecture for the unassisted capacity is false. In fact, it isn't even true for a channel with a single receiver~\cite{superadd-qcap}.

\subsection{Generalization to more receivers}
It is possible to generalize the protocol to more than two receivers. Without going into details, it is straightforward to show that a one-shot version of the protocol holds if there are more receivers; we simply get equations of the form of equations (\ref{eqn:oneshot-one}) and (\ref{eqn:oneshot-two}) for each receiver, and then we put them together in a way that is analogous to what we have done for two receivers.

To generalize this to the i.i.d. setting, the idea is to use a multiparty version of the FQSW protocol to decouple all the $A_1 \cdots A_n$ subsystems \cite{hayden-savov}. Thus, instead of simply having a constraint on $Q_1 + Q_2$, we get nontrivial constraints on every possible subset of receivers. The result is the following rate region:
\begin{equation}
    \sum_{j \in \mathcal{K}} Q_j \leqslant \frac{1}{2} \left[ \sum_{j \in \mathcal{K}} I(A_j;B_j) - J(A_{\mathcal{K}}) \right]
    \label{eqn:multiparty-iid}
\end{equation}
where $J(A_{\mathcal{K}}) = H(A_{j_1}) + \cdots + H(A_{j_{|\mathcal{K}|}}) - H(A_{j_1}\cdots A_{j_{|\mathcal{K}|}})$, for all $\mathcal{K} = \left\{ j_1, \cdots, j_{|\mathcal{K}|} \right\} \subseteq \{1,\cdots,m\}$. The mutual informations are defined on the state $\ket{\phi^{\mathcal{N}}}^{A_1\cdots A_n B_1 \cdots B_n D E} = U_{\mathcal{N}}\ket{\phi}^{A_1\cdots A_n A' D}$.

\section{Single-letter example}
In the classical case, the simplest example of a broadcast channel for which Marton's region is optimal is a deterministic channel, i.e. a channel where the outputs are completely determined by the inputs. Similarly, we can show that our rate region is optimal for entanglement-assisted quantum transmission through classical deterministic channels. This is perhaps unsurprising since entanglement would be highly unlikely to help classical transmission through a classical channel, but it nonetheless provides an example for which our theroem is optimal.

We say that $\mathcal{N}^{A' \rightarrow B_1 B_2}$ is a classical deterministic broadcast channel if there exist two deterministic functions $f_1:\{1,\ldots,|A'|\} \rightarrow \{1,\ldots,|B_1|\}$ and $f_2:\{1,\ldots,|A'|\} \rightarrow \{1,\ldots,|B_2|\}$ such that $U_{\mathcal{N}}\ket{i} = \ket{f_1(i)}^{B_1} \otimes \ket{f_2(i)}^{B_2} \otimes \ket{i}^E$ for some fixed orthonormal bases on $A'$, $B_1$, $B_2$ and $E$. We claim that any rate point that can be achieved for such a channel is a convex combination of rates which can be achieved via our coding method with input states of the form $\varphi^{A_1 A_2 A'} = \sum_{i = 1}^{|A'|} p_i \ketbra{f_1(i)}^{A_1} \otimes \ketbra{f_2(i)}^{A_2} \otimes \ketbra{i}^{A'}$ for some probability distribution $\{p_i\}$. To prove this, we first need the following observation:

\begin{lemma}\label{lem:single-letter}
	Let $f : \{1,\ldots,|D|\} \rightarrow \{1,\ldots,|B|\}$ be a function, and $\ket{\xi}^{ABCD}$ be $\sum_i \alpha_i \ket{\mu_i}^A \otimes \ket{f(i)}^B \otimes \ket{\nu_i}^C \otimes \ket{i}^D$, where $\ket{\mu_i}$ and $\ket{\nu_i}$ are any pure states, and $\ket{i}$ and $\ket{f(i)}$ represent $i$ and $f(i)$ encoded in a standard bases on $D$ and $B$ respectively. Then, $I(A;B)_{\xi} \leqslant H(B)_{\xi}$.
\end{lemma}
\begin{proof}
	The lemma simply follows from the observation that $\xi^{AB}$ is separable.
\end{proof}

Armed with this, we can now show the following:
\begin{thm}
	Let $\mathcal{N}^{A' \rightarrow B_1 B_2}$ be a classical deterministic channel. Then, the capacity region for this channel is the same as the achievable rate region given by Theorem \ref{thm:iid}.
\end{thm}
\begin{proof}
	According to the regularized converse theorem (Theorem \ref{thm:reg-converse}), for any achievable rate point $(Q_1,Q_2)$, there exists a state $\ket{\psi}^{A_1 A_2 B_1^n B_2^n E^n D} = U_{\mathcal{N}}^{\otimes n}\ket{\varphi}^{A_1 A_2 A'^n D}$ such that $Q_1 = \frac{1}{2n} I(A_1;B_1^n)_{\psi} + \delta_n$, $Q_2 = \frac{1}{2n} I(A_2;B_2^n)_{\psi} + \delta_n$, where $\delta_n \geqslant 0$, and $I(A_1;A_2)_{\psi} = 0$. Let $B_{1,i}$ and $B_{2,i}$ be the $i$th copies of $B_1$ and $B_2$ in $B_1^n$ and $B_2^n$, and, for each $i$, let $\psi_i^{A_1 A_2 B_1 B_2} = \sum_{jk} \ket{jkjk}\bra{jk} \psi^{B_{1,i} B_{2,i}} \ket{jk} \bra{jkjk}$, where the $\bra{jkjk}\ket{jk}$ are defined in the classical basis on $B_{1,i}$ and $B_{2,i}$ and in some fixed basis on $A_1, A_2, B_1$ and $B_2$. Then, we can bound the individual rates as follows:

\begin{align}
	Q_1 &\leqslant \frac{1}{2n} I(A_1;B_1^n)_{\psi} + \delta_n\\
	&\leqslant \frac{1}{2n} H(B_1^n)_{\psi} + \delta_n\\
	&\leqslant \frac{1}{2n} \sum_i H(B_{1,i})_{\psi} + \delta_n\\
	&= \frac{1}{2n} \sum_i H(B_{1})_{\psi_i} + \delta_n\\
	&= \frac{1}{n} \sum_i \frac{1}{2}I(A_1;B_1)_{\psi_i + \delta_n} \label{eqn:single-letter-terms}
\end{align}
and likewise for $Q_2$. The second inequality is due to Lemma \ref{lem:single-letter}, with the roles of the $B$ and $D$ subsystems in the lemma played by $B_1^n$ and $E^n$ respectively, and the third inequality makes use the subadditivity of the von Neumann entropy.

We can now do the same thing for the sum rate:
\begin{multline}
	Q_1 + Q_2\\
	\begin{split}
	&= \frac{1}{2n} \left\{ I(A_1;B_1^n)_{\psi} + I(A_2;B_2^n)_{\psi} \right\} + 2\delta_n\\
	&= \frac{1}{2n} \left\{ H(A_1)_{\psi} + H(A_2)_{\psi} - H(A_1|B_1^n)_{\psi} \right.\\
	&{} \hspace{0.5cm}\left. - H(A_1;B_2^n)_{\psi} \right\} + 2\delta_n\\
	&\leqslant \frac{1}{2n} \left\{ H(A_1A_2)_{\psi} - H(A_1A_2|B_1^nB_2^n)_{\psi} \right\} + 2\delta_n\\
	&= \frac{1}{2n} I(A_1A_2;B_1^nB_2^n)_{\psi} + 2\delta_n\\
	&\leqslant \frac{1}{2n} H(B_1^nB_2^n)_{\psi} + 2\delta_n\\
	&\leqslant \frac{1}{2n} \sum_i H(B_{1,i}B_{2,i})_{\psi} + 2\delta_n\\
	&= \frac{1}{2n} \sum_i H(B_{1}B_{2})_{\psi_i} + 2\delta_n\\
	&= \frac{1}{n} \sum_i \frac{1}{2} \left\{ H(B_1)_{\psi_i} + H(B_2)_{\psi_i} - I(B_1;B_2)_{\psi_i} \right\} + 2\delta_n\\
	&= \frac{1}{n} \sum_i \frac{1}{2} \left\{ I(A_1;B_1)_{\psi_i} + I(A_2;B_2)_{\psi_i} \right.\\
	&{} \hspace{0.5cm} \left. - I(A_1;A_2)_{\psi_i} \right\} + 2\delta_n \label{eqn:single-letter-terms-sumrate}
	\end{split}
\end{multline}
where, in the first inequality, we have made use of the fact that $A_1$ and $A_2$ are independent and of the standard inequality $H(AB|CD) \leqslant H(A|C) + H(B|D)$, and the last equality follows from the special form of the $\psi_i$'s.

Since every $i$ in equations (\ref{eqn:single-letter-terms}) and (\ref{eqn:single-letter-terms-sumrate}) corresponds to a rate which is achievable via Theorem \ref{thm:iid}, this concludes the proof.
\end{proof}

\section{Discussion}\label{sec:discussion}
We have shown that a new protocol for entanglement-assisted communication of quantum information through quantum broadcast channels can be obtained from the FQSW protocol. Our protocol achieves the following rate region for every state $\ket{\phi}^{A_1 A_2 A' D}$:
\begin{equation}
\begin{split}
0 \leqslant Q_1 &\leqslant \frac{1}{2} I(A_1; B_1)_\psi\\
0 \leqslant Q_2 &\leqslant \frac{1}{2} I(A_2; B_2)_\psi\\
Q_1 + Q_2 &\leqslant \frac{1}{2} \left[ I(A_1;B_1)_\psi + I(A_2;B_2)_\psi - I(A_1;A_2)_\psi \right].
\end{split}
\end{equation}
where $\ket{\psi}^{A_1 A_2 B_1 B_2 D E} = U_{\mathcal{N}}^{A' \rightarrow B_1 B_2 E}\ket{\phi}^{A_1 A_2 A' D}$.

The corresponding rate region (equation (\ref{eqn:iidregion})) is very similar to Marton's region for classical broadcast channels (equation (\ref{eqn:marton})) \cite{marton}; except for the factors of $1/2$, the two expressions are identical. In fact, for classical channels, the rates for entanglement-assisted quantum communication found here can be achieved directly using teleportation between the senders and the receiver, with the classical communication required by teleportation transmitted using Marton's protocol. From this point of view, our results can be viewed as a direct generalization of Marton's region to quantum channels.

Therefore, once again, it is the entanglement-assisted version of the quantum capacity that bears the strongest resemblance to its classical counterpart. The same is true for both the regular point-to-point quantum channel \cite{BSST02} and the quantum multiple-access channel \cite{hdw05} \cite{merging}. In both those cases, the known achievable rate regions for entanglement-assisted quantum communication are identical to their classical counterparts.  This collection of similarities suggests a fundamental question. To what extent does the addition of free entanglement make quantum information theory similar to classical information theory? 

Of course, the lack of a single-letter converse for Marton's region and, by extension, for our region, leaves open the possibility that the analogy might break down for a new, better broadcast region that remains to be discovered. A first step towards eliminating that uncertainty could be to find a better characterization of the quantum regions we have presented here. The presence of the ``discarded'' system $D$ in Theorem \ref{thm:iid} is equivalent to optimizing over all mixed states $\phi^{A_1 A_2 A'}$ rather than only over pure states. This is not required for most theorems in quantum information theory, but we have not found a way to prove the regularized converse without allowing for the possibility of mixed states. We leave it as an open problem to determine whether it is possible to demonstrate a converse theorem that does not require allowing mixed states.


Finally, for the unassisted case, it is very interesting to note the absence of an independent constraint on the sum-rate. However, we already know that this region is suboptimal even for channels with a single receiver. It would therefore be desirable to know whether this holds for the true capacity region and whether there is an underlying principle which explains this phenomenon.

\subsection*{Acknowledgments}
The authors would like to thank Gilles Brassard, Igor Devetak, Young-Han Kim, Ivan Savov, Andreas Winter and Jon Yard for conversations that helped them in this research. They are also grateful for support from CIAR, the Canada Research Chairs program, FQRNT, MITACS and NSERC.

\appendices

\section{Asymptotic equality}\label{sec:asymptotic}
Here we formally define asymptotic equality denoted by the symbol $\aeq$. Let $\psi = \left\{ \psi_{(1)},\psi_{(2),\cdots} \right\}$ and $\varphi = \left\{ \varphi_{(1)},\varphi_{(2),\cdots} \right\}$ be two families of quantum states, where $\psi_{(n)}$ and $\varphi_{(n)}$ are defined on a Hilbert space $\mathcal{H}^{\otimes n}$. Then we say that $\psi \aeq \varphi$ if $\lim_{n \rightarrow \infty} \left\| \psi_{(n)} - \varphi_{(n)} \right\|_1 = 0$. We then say that $\psi$ and $\varphi$ are asymptotically equal. Note that, by the triangle inequality, $\aeq$ is transitive for any finite number of steps independent of $n$.

It should be mentioned that throughout the paper, asymptotic families of states are not always explicitly referred to as such, but generally speaking, whenever a state depends on the number of copies, it should be considered as a family of states. In addition, with a slight abuse of notation, we allow quantum operations on families of states; it should be clear which operation is done on each member of the family.

\section{Typical subspaces}\label{sec:typical}
Much of information theory relies on the concept of typical sequences. Let $\mathcal{X}$ be some alphabet and let $X$ be a random variable defined on $\mathcal{X}$ and distributed according to $p(x)$. Define the $\varepsilon$-typical set as follows:
\begin{equation*}
   \mathcal{T}_{\varepsilon}^{(n)} = \left\{ x^n \in \mathcal{X}^n \Big| \left|{-\tfrac{1}{n}}\log \Pr\{X^n = x^n\} - H(X) \right| \leqslant \varepsilon \right\}
\end{equation*}
where $X^n$ refers to $n$ independent, identically-distributed copies of $X$. It can be shown that the two following properties hold:
\begin{enumerate}
    \item There exists a function $\varepsilon(n)$ such that $\lim_{n \rightarrow \infty} \varepsilon(n) = 0$ and such that $\Pr\{X^n \in \mathcal{T}_{\varepsilon(n)}^{(n)}\} \geqslant 1-\varepsilon(n)$.
    \item There exists an $n_0$ such that for all $n>n_0$, $|\mathcal{T}_{\varepsilon}^{(n)}| \leqslant 2^{n[H(X) + \varepsilon]}$.
\end{enumerate}

The quantum generalization of these concepts is relatively straightforward: let $\rho^A = \sum_{x \in \mathcal{X}} p(x) \ketbra{x}$ be the spectral decomposition of a quantum state $\rho^A$ on a quantum system $A$. Then we can define the typical projector on the quantum system $A^{n}$ as follows:
\begin{equation*}
    \Pi_{\varepsilon}^{(n)} = \sum_{x^n \in \mathcal{T}_{\varepsilon}^{(n)}} \ketbra{x^n}
\end{equation*}
We call the support of $\Pi_{\varepsilon}^{(n)}$ the $\varepsilon$-typical subspace of $A^{n}$. (For brevity, we often omit $\varepsilon$ and refer simply to the typical subspace. In this case, unless otherwise stated, $\varepsilon$ can be assumed to be a positive constant, independent of $n$.) The two properties given above generalize to the quantum case:
\begin{enumerate}
    \item There exists a function $\varepsilon(n)$ such that $\lim_{n \rightarrow \infty} \varepsilon(n) = 0$ and such that $\tr\left[\Pi_{\varepsilon(n)}^{(n)} {\rho^A}^{\otimes n}\right] \geqslant 1-\varepsilon(n)$.
    \item There exists an $n_0$ such that for all $n>n_0$, $\tr[\Pi_{\varepsilon}^{(n)}] \leqslant 2^{n[H(A) + \varepsilon]}$.
\end{enumerate}

Note that the first of these two properties implies that $\Pi_{\varepsilon(n)}^{(n)} \cdot {\rho^A}^{\otimes n} \aeq {\rho^A}^{\otimes n}$, via the ``gentle measurement'' lemma (Lemma 9 in \cite{winter99}). One can also easily show that the normalized version of $\Pi_{\varepsilon(n)}^{(n)} \cdot {\rho^A}^{\otimes n}$ is also asymptotically equal to ${\rho^A}^{\otimes n}$, and that it also holds for i.i.d. states with more than one subsystem.

\section{Uhlmann's theorem} \label{sec:uhlmann}
In this paper, we use Uhlmann's theorem \cite{uhlmann} several times, in the form first presented as Lemma 2.2 in \cite{DHW05}:
\begin{thm}
    Let $\ket{\psi}^{AB}$ and $\ket{\varphi}^{AB'}$ be two quantum states such that $\left\| \psi^A - \varphi^A \right\|_1 \leqslant \varepsilon$. Then there exists an isometry $U^{B' \rightarrow B}$ such that $\left\| \psi^{AB} - U^{B' \rightarrow B} \cdot \varphi^{AB'} \right\|_1 \leqslant 2 \sqrt{\varepsilon}$.
\end{thm}

\bibliographystyle{IEEEtran}
\bibliography{IEEEabrv,broadcast}

\end{document}

%% file: quantum.tex
\usepackage{amsmath}
\usepackage{amssymb}

\newcommand{\bra}[1]{\langle #1|}
\newcommand{\ket}[1]{|#1 \rangle}

\newcommand{\ketbra}[1]{\ket{#1}\bra{#1}}

\newcommand{\ident}{\mathbb{I}}
\DeclareMathOperator{\tr}{Tr}

%% file: achievable.pdf_t
\begin{picture}(0,0)%
\includegraphics{achievable.pdf}%
\end{picture}%
\setlength{\unitlength}{3947sp}%
\begingroup\makeatletter\ifx\SetFigFontNFSS\undefined%
\gdef\SetFigFontNFSS#1#2#3#4#5{%
  \reset@font\fontsize{#1}{#2pt}%
  \fontfamily{#3}\fontseries{#4}\fontshape{#5}%
  \selectfont}%
\fi\endgroup%
\begin{picture}(6399,5511)(64,-4789)
\put(226,389){\makebox(0,0)[lb]{\smash{{\SetFigFontNFSS{12}{14.4}{\rmdefault}{\mddefault}{\updefault}{\color[rgb]{0,0,0}Reference 1}%
}}}}
\put(2551,-2161){\makebox(0,0)[lb]{\smash{{\SetFigFontNFSS{12}{14.4}{\rmdefault}{\mddefault}{\updefault}{\color[rgb]{0,0,0}$W$}%
}}}}
\put(4576,-1786){\makebox(0,0)[lb]{\smash{{\SetFigFontNFSS{12}{14.4}{\rmdefault}{\mddefault}{\updefault}{\color[rgb]{0,0,0}$E$}%
}}}}
\put(226,-361){\makebox(0,0)[lb]{\smash{{\SetFigFontNFSS{12}{14.4}{\familydefault}{\mddefault}{\updefault}{\color[rgb]{0,0,0}Bob 1}%
}}}}
\put(151,-2086){\makebox(0,0)[lb]{\smash{{\SetFigFontNFSS{12}{14.4}{\familydefault}{\mddefault}{\updefault}{\color[rgb]{0,0,0}Alice}%
}}}}
\put(226,-3586){\makebox(0,0)[lb]{\smash{{\SetFigFontNFSS{12}{14.4}{\familydefault}{\mddefault}{\updefault}{\color[rgb]{0,0,0}Bob 2}%
}}}}
\put(2026,-3586){\makebox(0,0)[lb]{\smash{{\SetFigFontNFSS{12}{14.4}{\familydefault}{\mddefault}{\updefault}{\color[rgb]{0,0,0}$\widetilde{B_2}$}%
}}}}
\put(2026,-2986){\makebox(0,0)[lb]{\smash{{\SetFigFontNFSS{12}{14.4}{\familydefault}{\mddefault}{\updefault}{\color[rgb]{0,0,0}$\widetilde{A_2}$}%
}}}}
\put(2026,-2461){\makebox(0,0)[lb]{\smash{{\SetFigFontNFSS{12}{14.4}{\familydefault}{\mddefault}{\updefault}{\color[rgb]{0,0,0}$A_2$}%
}}}}
\put(2026,-1936){\makebox(0,0)[lb]{\smash{{\SetFigFontNFSS{12}{14.4}{\familydefault}{\mddefault}{\updefault}{\color[rgb]{0,0,0}$A_1$}%
}}}}
\put(2026,-1261){\makebox(0,0)[lb]{\smash{{\SetFigFontNFSS{12}{14.4}{\familydefault}{\mddefault}{\updefault}{\color[rgb]{0,0,0}$\widetilde{A_1}$}%
}}}}
\put(2026,-586){\makebox(0,0)[lb]{\smash{{\SetFigFontNFSS{12}{14.4}{\familydefault}{\mddefault}{\updefault}{\color[rgb]{0,0,0}$\widetilde{B_1}$}%
}}}}
\put(2026,539){\makebox(0,0)[lb]{\smash{{\SetFigFontNFSS{12}{14.4}{\familydefault}{\mddefault}{\updefault}{\color[rgb]{0,0,0}$R_1$}%
}}}}
\put(3001,-1861){\makebox(0,0)[lb]{\smash{{\SetFigFontNFSS{12}{14.4}{\familydefault}{\mddefault}{\updefault}{\color[rgb]{0,0,0}$A'$}%
}}}}
\put(3151,-2986){\makebox(0,0)[lb]{\smash{{\SetFigFontNFSS{12}{14.4}{\familydefault}{\mddefault}{\updefault}{\color[rgb]{0,0,0}$\widehat{A}$}%
}}}}
\put(4051,-3136){\makebox(0,0)[lb]{\smash{{\SetFigFontNFSS{12}{14.4}{\familydefault}{\mddefault}{\updefault}{\color[rgb]{0,0,0}$B_2$}%
}}}}
\put(4051,-1111){\makebox(0,0)[lb]{\smash{{\SetFigFontNFSS{12}{14.4}{\familydefault}{\mddefault}{\updefault}{\color[rgb]{0,0,0}$B_1$}%
}}}}
\put(5476,-511){\makebox(0,0)[lb]{\smash{{\SetFigFontNFSS{12}{14.4}{\familydefault}{\mddefault}{\updefault}{\color[rgb]{0,0,0}$V_1$}%
}}}}
\put(5476,-3661){\makebox(0,0)[lb]{\smash{{\SetFigFontNFSS{12}{14.4}{\familydefault}{\mddefault}{\updefault}{\color[rgb]{0,0,0}$V_2$}%
}}}}
\put(6301,-3511){\makebox(0,0)[lb]{\smash{{\SetFigFontNFSS{12}{14.4}{\familydefault}{\mddefault}{\updefault}{\color[rgb]{0,0,0}$\widehat{B_2}$}%
}}}}
\put(6301,-3961){\makebox(0,0)[lb]{\smash{{\SetFigFontNFSS{12}{14.4}{\familydefault}{\mddefault}{\updefault}{\color[rgb]{0,0,0}$\bar{B}_2$}%
}}}}
\put(6376,-286){\makebox(0,0)[lb]{\smash{{\SetFigFontNFSS{12}{14.4}{\familydefault}{\mddefault}{\updefault}{\color[rgb]{0,0,0}$\bar{B}_1$}%
}}}}
\put(6376,-661){\makebox(0,0)[lb]{\smash{{\SetFigFontNFSS{12}{14.4}{\familydefault}{\mddefault}{\updefault}{\color[rgb]{0,0,0}$\widehat{B_1}$}%
}}}}
\put(226,-4711){\makebox(0,0)[lb]{\smash{{\SetFigFontNFSS{12}{14.4}{\familydefault}{\mddefault}{\updefault}{\color[rgb]{0,0,0}Reference 2}%
}}}}
\put(2026,-4711){\makebox(0,0)[lb]{\smash{{\SetFigFontNFSS{12}{14.4}{\familydefault}{\mddefault}{\updefault}{\color[rgb]{0,0,0}$R_2$}%
}}}}
\put(3601,-2086){\makebox(0,0)[lb]{\smash{{\SetFigFontNFSS{12}{14.4}{\familydefault}{\mddefault}{\updefault}{\color[rgb]{0,0,0}$U_{\mathcal{N}}$}%
}}}}
\end{picture}%

%% file: bigbroadcast.pdf_t
\begin{picture}(0,0)%
\includegraphics{bigbroadcast.pdf}%
\end{picture}%
\setlength{\unitlength}{3947sp}%
\begingroup\makeatletter\ifx\SetFigFontNFSS\undefined%
\gdef\SetFigFontNFSS#1#2#3#4#5{%
  \reset@font\fontsize{#1}{#2pt}%
  \fontfamily{#3}\fontseries{#4}\fontshape{#5}%
  \selectfont}%
\fi\endgroup%
\begin{picture}(6399,4611)(64,-3889)
\put(226,389){\makebox(0,0)[lb]{\smash{{\SetFigFontNFSS{12}{14.4}{\rmdefault}{\mddefault}{\updefault}{\color[rgb]{0,0,0}Reference 1}%
}}}}
\put(2026,539){\makebox(0,0)[lb]{\smash{{\SetFigFontNFSS{12}{14.4}{\rmdefault}{\mddefault}{\updefault}{\color[rgb]{0,0,0}$R_1$}%
}}}}
\put(4876,-2686){\makebox(0,0)[lb]{\smash{{\SetFigFontNFSS{12}{14.4}{\rmdefault}{\mddefault}{\updefault}{\color[rgb]{0,0,0}$B_2$}%
}}}}
\put(226,-136){\makebox(0,0)[lb]{\smash{{\SetFigFontNFSS{12}{14.4}{\familydefault}{\mddefault}{\updefault}{\color[rgb]{0,0,0}Bob 1}%
}}}}
\put(226,-1636){\makebox(0,0)[lb]{\smash{{\SetFigFontNFSS{12}{14.4}{\familydefault}{\mddefault}{\updefault}{\color[rgb]{0,0,0}Alice}%
}}}}
\put(226,-3136){\makebox(0,0)[lb]{\smash{{\SetFigFontNFSS{12}{14.4}{\familydefault}{\mddefault}{\updefault}{\color[rgb]{0,0,0}Bob 2}%
}}}}
\put(226,-3811){\makebox(0,0)[lb]{\smash{{\SetFigFontNFSS{12}{14.4}{\familydefault}{\mddefault}{\updefault}{\color[rgb]{0,0,0}Reference 2}%
}}}}
\put(2026,-211){\makebox(0,0)[lb]{\smash{{\SetFigFontNFSS{12}{14.4}{\familydefault}{\mddefault}{\updefault}{\color[rgb]{0,0,0}$\widetilde{B_1}$}%
}}}}
\put(2026,-811){\makebox(0,0)[lb]{\smash{{\SetFigFontNFSS{12}{14.4}{\familydefault}{\mddefault}{\updefault}{\color[rgb]{0,0,0}$\widetilde{A_1}$}%
}}}}
\put(2026,-1261){\makebox(0,0)[lb]{\smash{{\SetFigFontNFSS{12}{14.4}{\familydefault}{\mddefault}{\updefault}{\color[rgb]{0,0,0}$A_1$}%
}}}}
\put(2026,-2011){\makebox(0,0)[lb]{\smash{{\SetFigFontNFSS{12}{14.4}{\familydefault}{\mddefault}{\updefault}{\color[rgb]{0,0,0}$A_2$}%
}}}}
\put(2026,-2536){\makebox(0,0)[lb]{\smash{{\SetFigFontNFSS{12}{14.4}{\familydefault}{\mddefault}{\updefault}{\color[rgb]{0,0,0}$\widetilde{A_2}$}%
}}}}
\put(2026,-3136){\makebox(0,0)[lb]{\smash{{\SetFigFontNFSS{12}{14.4}{\familydefault}{\mddefault}{\updefault}{\color[rgb]{0,0,0}$\widetilde{B_2}$}%
}}}}
\put(2026,-3811){\makebox(0,0)[lb]{\smash{{\SetFigFontNFSS{12}{14.4}{\familydefault}{\mddefault}{\updefault}{\color[rgb]{0,0,0}$R_2$}%
}}}}
\put(4876,-661){\makebox(0,0)[lb]{\smash{{\SetFigFontNFSS{12}{14.4}{\familydefault}{\mddefault}{\updefault}{\color[rgb]{0,0,0}$B_1$}%
}}}}
\put(5401,-1336){\makebox(0,0)[lb]{\smash{{\SetFigFontNFSS{12}{14.4}{\familydefault}{\mddefault}{\updefault}{\color[rgb]{0,0,0}$E$}%
}}}}
\put(4276,-1636){\makebox(0,0)[lb]{\smash{{\SetFigFontNFSS{12}{14.4}{\familydefault}{\mddefault}{\updefault}{\color[rgb]{0,0,0}$U_{\mathcal{N}}$}%
}}}}
\put(5626,-2161){\makebox(0,0)[lb]{\smash{{\SetFigFontNFSS{12}{14.4}{\familydefault}{\mddefault}{\updefault}{\color[rgb]{0,0,0}$\widehat{A}$}%
}}}}
\put(5626,-2386){\makebox(0,0)[lb]{\smash{{\SetFigFontNFSS{12}{14.4}{\familydefault}{\mddefault}{\updefault}{\color[rgb]{0,0,0}$\breve{A}_1$}%
}}}}
\put(2926,-1711){\makebox(0,0)[lb]{\smash{{\SetFigFontNFSS{12}{14.4}{\familydefault}{\mddefault}{\updefault}{\color[rgb]{0,0,0}$W$}%
}}}}
\put(3451,-1411){\makebox(0,0)[lb]{\smash{{\SetFigFontNFSS{12}{14.4}{\familydefault}{\mddefault}{\updefault}{\color[rgb]{0,0,0}$A'$}%
}}}}
\end{picture}%